\documentclass[namedreferences]{SolarPhysics}
\usepackage[optionalrh]{spr-sola-addons} 
\usepackage{epsfig}                     
\usepackage{graphicx}                    
\usepackage{color}                       
\usepackage{url}                         

\newcommand{\adv}{    {\it Adv. Space Res.}}

\newcommand{\apj}{    {\it Astrophys. J.}}

\newcommand{\grl}{    {\it Geophys. Res. Lett.}}

\newcommand{\jastp}{  {\it J. Atmos. Solar Terr. Phys.}}
\newcommand{\jgr}{    {\it J. Geophys. Res.}}

\newcommand{\nat}{    {\it Nature}}

\newcommand{\solphys}{{\it Solar Phys.}}

\newcommand{\ssr}{    {\it Space Sci. Rev.}}

\begin{document}

\begin{article}

\begin{opening}

\title{The Variation of Solar Wind Correlation Lengths Over Three Solar Cycles}

%
\author{R. T. ~\surname{Wicks}$^{1}$\sep M. J. ~\surname{Owens}$^{1}$\sep T. S. ~\surname{Horbury}$^{1}$}

%
\runningauthor{Wicks et al.}
\runningtitle{The Variation of Solar Wind Correlation Lengths Over Three Solar Cycles}

%
  \institute{$^{1}$ Space and Atmospheric Physics Group, Department of Physics, Imperial College London, UK.
                     email: \url{r.wicks@imperial.ac.uk}}


\begin{abstract}

We present the results of a study of solar wind velocity and magnetic field correlation lengths over the last 35 years. The correlation length of the magnetic field magnitude $\lambda_{|B|}$ increases on average by a factor of two at solar maxima compared to solar minima. The correlation lengths of the components of the magnetic field $\lambda_{B_{XYZ}}$ and of the velocity $\lambda_{V_{YZ}}$ do not show this change and have similar values, indicating a continual turbulent correlation length of around $1.4 \times 10^6$ km. We conclude that a linear relation between $\lambda_{|B|}$, $VB^2$ and Kp suggests that the former is related to the total magnetic energy in the solar wind and an estimate of the average size of geo-effective structures which is in turn proportional to $VB^2$. By looking at the distribution of daily correlation lengths we show that the solar minimum values of $\lambda_{|B|}$ correspond to the turbulent outer scale. A tail of larger $\lambda_{|B|}$ values is present at solar maximum, causing the increase in mean value.

\end{abstract}

%

\end{opening}

\section{Introduction}

The 22-year solar magnetic polarity cycle has been observed as the approximately 11-year sunspot activity cycle for at least the last 250 years \cite{Mursula01}. The effects of this cycle can be seen throughout the heliosphere: in the behaviour of the solar wind \cite{Richardson08}, the flux of cosmic rays measured on Earth \cite{Forbush54}, in auroral indices \cite{Silverman92}, geomagnetic activity \cite{Cliver96}, the nonlinearity of the response of the magnetosphere to the solar wind \cite{Johnson05}, and in the rate of coronal mass ejections (CMEs) and hence the average $|B|$ at Earth \cite{Owens06}.
\par
The interaction of solar particles and magnetic field with the rest of the heliosphere is mediated by the solar wind. The solar wind magnetic field and velocity fluctuations are turbulent \cite{Goldstein95} and the solar wind also contains many different forms of large scale structure, such as CMEs \cite{Wimmer-Schweingruber06} and co-rotating interaction regions (CIRs) formed by interacting streams with different velocities \cite{Gosling99}. Turbulence and these larger structures affect the scattering of cosmic rays \cite{Minnie03} and the interaction of the solar wind with planetary magnetospheres, causing auroral activity \cite{Legrand85}.
\par
One method for studying turbulent fluids is to look at the turbulent correlation length $\lambda$ \cite{Matthaeus94}. This is the average distance over which correlation persists in the flow and is normally estimated from a single time series using the Taylor hypothesis \cite{Taylor}. Direct measurements of the correlation length have been made using multiple spacecraft \cite{Matthaeus05, Wicks09}, however they do not differ significantly from those made using single time series. Correlation length in the solar wind affects the scattering of energetic particles \cite{Parhi02} such as cosmic rays \cite{Minnie03}, it is an important parameter of models of the heliosphere \cite{Matthaeus94}, giving an appropriate scale size for the models to resolve, and can be considered as a typical size for energy containing structures interacting with planetary magnetospheres.
\par
Given that the solar wind correlation length and the solar activity cycle both affect the scattering of cosmic rays and the behaviour of the Earth's magnetosphere it seems appropriate to study the solar cycle variation of the correlation length, its causes and its effects. In situ observations of the solar wind around 1 AU have been made regularly since the early 1970s using a variety of different spacecraft. Here we use the IMP8 and ISEE3 spacecraft, and the multi-spacecraft OMNI one minute resolution data set to provide magnetic field and velocity observations from 1974 until 2009. This data set covers the last three solar cycles and so allows the current unusual solar minimum \cite{McComas08, Owens08} to be compared to previous minima. We compare the correlation lengths to the changing sunspot number, CME rate, and magnetospheric activity as measured by the daily total of the Kp index over the most recent solar cycle, as well as looking at the general trends in correlation lengths for the last 35 years.

\section{Long term trends}

The data used solar wind velocity $V$ and magnetic field $B$ vectors and magnitudes, all are one minute resolution. All observations are made close to 1 AU distance from the Sun. Here we use these data to provide a daily estimate of the correlation length of the velocity and magnetic field in the solar wind on time scales shorter than a day over the last three solar cycles.
\par
Correlation lengths are estimated by using up to 960 consecutive one minute cadence data points to calculate the autocorrelation function of the data, lagged by a time $\tau$ between 0 and 480 minutes, adding up to a total of 1440 minutes, allowing the calculation of a correlation length every day. An autocorrelation curve $C(\tau)$ is calculated in this way every day and the time lag multiplied by the average solar wind speed is used to convert the time scale into a radial spatial scale, that is $C(x) = |V| C(\tau)$. The resulting spatial autocorrelation curve is a measure of the correlation of any scalar quantity (component of a vector) in the radial direction. This curve is then interpolated onto a standard spatial scale between $0$ and $9 \times 10^6$ km with a resolution of $3 \times 10^4$ km. This allows spatial correlation curves from different days to be compared directly. Finally a correlation length $\lambda$ is estimated from the correlation curves by fitting an exponential \begin{equation}
C(x) = exp(-\frac{x}{\lambda}).
\label{eq:Exp1}
\end{equation}
\par
Days are only accepted if the spacecraft is in the solar wind and if there are over 480 out of 960 observations available. Individual days have very different correlation properties, caused by the very variable structure seen in the solar wind on timescales of a day such as CMEs, large amplitude Alfv\'{e}n waves and CIRs.
\par

\begin{figure}
\centerline{\includegraphics[width=0.6\textwidth]{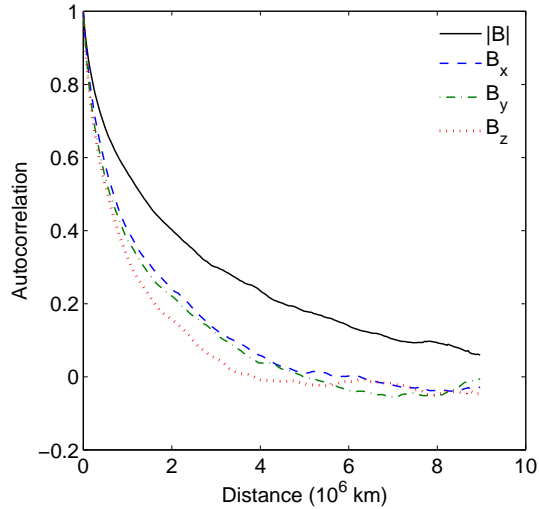}}
\caption{An example of the quarterly averaged spatial autocorrelation curves calculated for the solar wind magnetic field magnitude (solid line), GSE x component (dashed line), y component (dot-dashed line) and z component (dotted line).}
\label{fig:Quarterly_Correlation_Scale}
\end{figure}

To observe the change in $\lambda$ with solar cycle the daily autocorrelation functions are averaged over three months, giving a quarterly averaged correlation curve $\tilde{C}(x)$ from which an estimate of $\lambda$ is made. An example of the resulting mean autocorrelation curves are shown in Figure \ref{fig:Quarterly_Correlation_Scale}. Quarterly averages are made only if there are at least 10 suitable days in the quarter of a year. The conditions on number of data points per day and number of good days per quarter are made to remove the statistically worst data points and keep a minimum standard between the correlation lengths calculated using different spacecraft despite the different orbits, data gaps and data qualities that occur. The IMP8 spacecraft had a geocentric orbit and so spent approximately half of its time in the solar wind. The OMNI data, however, are almost continuous, having very few missing days and fewer data gaps within individual days. Thus the quarterly correlation lengths since 1995 calculated from the OMNI data provide the most complete record of correlation over a solar cycle. The ISEE3 spacecraft data are used only during its primary mission (1979 - June 1982) before it was renamed ICE and left near-Earth space for comet Giacobini-Zinner. We have used the ISEE3 data specifically to fill in a period of time when the IMP8 data are particularly sparse.
\par

\begin{figure}
\centerline{\includegraphics[width=0.75\textwidth]{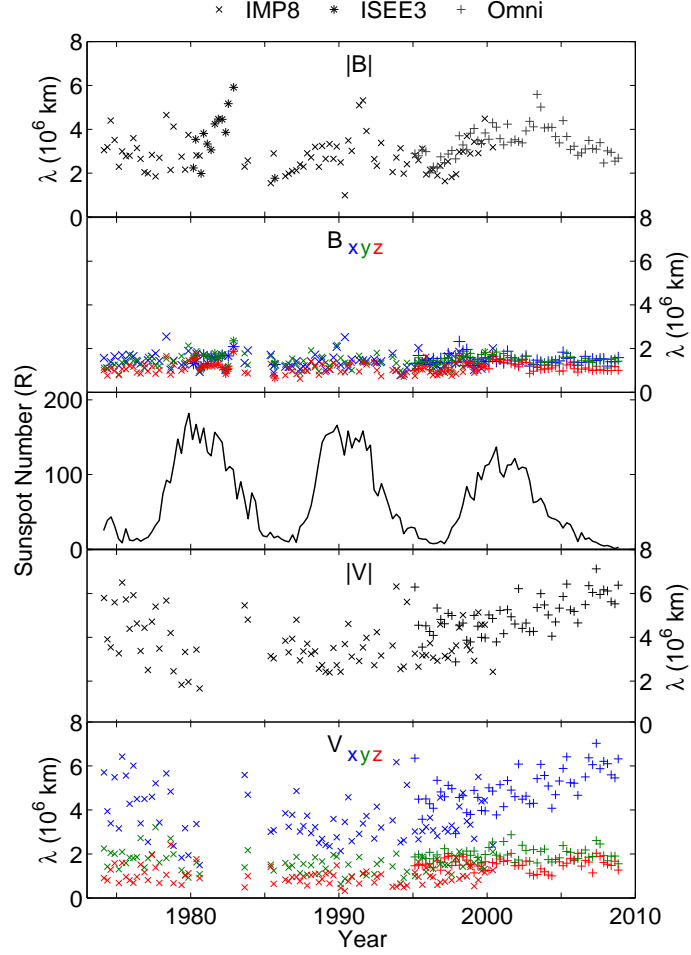}}
\caption{The quarterly averaged solar wind correlation length from 1974 until 2009. The top panel plots $\lambda_{|B|}$, the second panel $\lambda_{B_{XYZ}}$. The third panel shows \textbf{quarterly} averaged sunspot number (R) as an indicator of solar activity. The fourth panel plots $\lambda_{|V|}$. The bottom panel shows the components of velocity $\lambda_{V_{XYZ}}$, $\lambda_{V_X}$ is almost identical to $\lambda_{|V|}$.}
\label{fig:Three_Solar_Cycles}
\end{figure}

The quarterly correlation lengths are plotted in Figure \ref{fig:Three_Solar_Cycles}, the top two panels show $\lambda$ calculated from the magnetic field magnitude $\lambda_{|B|}$ and components $\lambda_{B_{XYZ}}$, the middle panel shows \textbf{quarterly} averaged sunspot number as an indicator of solar activity, and the bottom two panels show $\lambda$ calculated from the solar wind speed $\lambda_{|V|}$ and velocity components $\lambda_{V_{XYZ}}$. The errors on the estimations of $\lambda$ from these results are an order of magnitude smaller than the values themselves and hence are not plotted. Where the IMP8, ISEE3 and OMNI data sets overlap in time the correlation lengths calculated are similar in value. In particular between 1994 and 2001 both the IMP8 and OMNI data are available and although there is some systematic effect making the Omni correlation lengths slightly larger on average (particularly in the velocity measurements) they are not outside the scatter of the IMP8 results. This suggests that the estimates of the correlation length are consistent across all the data. 
\par
\textbf{We can also compare our values for correlation lengths to previous estimates. Single spacecraft studies at 1 AU typically produce values of approximately $\lambda_{B} = 1.12\times10^{6}$ km and $\lambda_V = 2.83\times10^{6}$ km \cite{Matthaeus82}. More recently it has become possible to use multiple spacecraft to calculate truly spatial correlations and the typical lengths of these are $\lambda_{B} = 1.19\times10^{6}$ km \cite{Matthaeus05}. Wicks et al., (2009) made the first study of the change of multi point correlation length measurements with solar cycle and found that at solar minimum $\lambda_{|B|} = 0.75\times10^{6}$ km and at solar maximum $\lambda_{|B|} = 1.75\times10^{6}$ km, but that the components of $\textbf{B}$ had a roughly constant correlation length of $\lambda_{B_{XYZ}} \approx 1.1\times10^{6}$ km. These results, made using many different techniques, are broadly consistent with those shown here and make us confident that our results are reliable.}
\par
The top panel of Figure \ref{fig:Three_Solar_Cycles} shows that $\lambda_{|B|}$ changes with the solar cycle. The four activity minima contained in the data all give rise to similar correlation lengths of $\lambda_{|B|} \approx 2 \times 10^6$ km. For each of the last three solar activity maxima there is a peak in correlation length shortly after the peak in sunspot number, although the peak of solar cycle 21 is partially missing due to data gaps. The change in $\lambda_{|B|}$ between minimum and maximum is approximately a factor of two in all three cycles and the very highest values of $\lambda_{|B|}$ in each cycle are over three times the minimum values. The most recent estimates of $\lambda_{|B|}$ made here are from the end of 2008 and have not yet reached the previous minimum value and so it seems that solar minimum had not yet been reached by the end of 2008, based on this measure.
\par
The second panel in Figure \ref{fig:Three_Solar_Cycles} plots the quarterly correlation lengths of the three GSE components of \textbf{B}. All three components have similar correlation lengths and remain approximately constant with average values of $\langle\lambda_{B_X}\rangle = 1.4 \pm 0.3 \times 10^6$ km, $\langle\lambda_{B_Y}\rangle = 1.4 \pm 0.3 \times 10^6$ km and $\langle\lambda_{B_Z}\rangle = 1.1 \pm 0.2 \times 10^6$ km over the whole 35 year time period. 
\par
The quarterly correlation lengths of solar wind speed $|V|$ and the GSE $V_x$ component shown in the bottom two panels of Figure \ref{fig:Three_Solar_Cycles} are almost identical, as one would expect. The only clear trend in the behaviour of $\lambda_{|V|}$ is an increase in value since the most recent solar maximum. The previous highest values are all at the beginning of activity minima (1975, 1984, 1994), particularly around 1975. The previous high values of $\lambda_{|V|}$ from 1974 until 1980 could indicate that the current behaviour of the solar wind velocity is in fact a return to longer correlation lengths which previously occurred in the weaker cycle 20. However, without earlier observations than used here it is impossible to be certain. 
\par
The correlation lengths of the GSE y and z components of \textbf{V} are plotted in the bottom panel of Figure \ref{fig:Three_Solar_Cycles}; like the components of \textbf{B} they remain approximately constant for the whole 35 year period, with average values of $\langle\lambda_{V_Y}\rangle = 1.7 \pm 0.4 \times 10^6$ km and $\langle\lambda_{V_Z}\rangle = 1.3 \pm 0.5 \times 10^6$ km. The components of \textbf{B} and \textbf{V} (with the exception of $V_x$ since it is effectively $|V|$) have very similar correlation lengths and do not vary measurably with time. Thus the perpendicular components of \textbf{V} and all the components of \textbf{B} have similar correlation lengths. $\lambda_{|B|}$ at solar minimum is also approximately the same size. This is consistent with the presence of continual, developed Alfv\'{e}nic turbulence at 1 AU in the ecliptic, that is unaffected by solar activity. Further implications are discussed below.

\section{Solar Cycle 23 in detail}

\begin{figure}
\centerline{\includegraphics[width=0.75\textwidth]{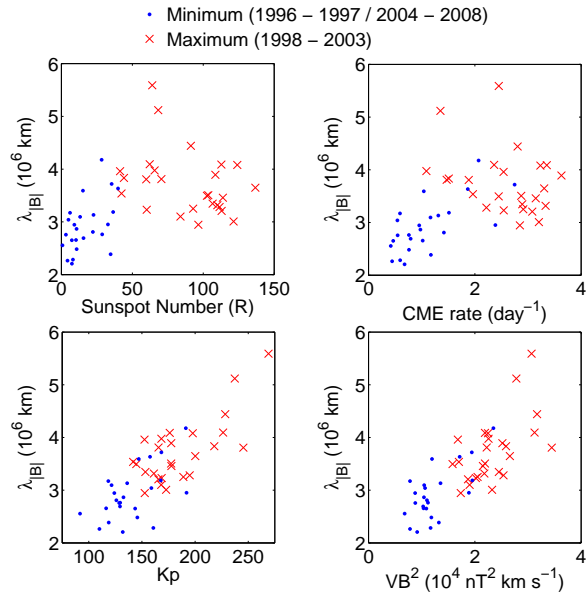}}
\caption{Plots of $\lambda_{|B|}$ during cycle 23 against Sunspot number (top left), average daily CME rate (top right), magnetospheric activity index Kp (bottom left) and solar wind quantity $VB^2$. The \textbf{minimum} phase of the solar cycle is plotted as $\bullet$ and the \textbf{maximum} phase as $\times$ to show the different behaviour during the different halves of the cycle. \textbf{Maximum is defined as when the sunspot number is larger than 40}.}
\label{fig:Solar_Cycle_23_L_vs_R_vs_Kp_vs_CMEs}
\end{figure}

Solar cycle 23 has the most complete data coverage of any of the three cycles studied here thus we investigate it further to identify the wider effects of the changing correlation length of $|B|$. To do this we compare the quarterly average $\lambda_{|B|}$ with sunspot number, estimated CME rate, mean daily total magnetospheric activity index Kp, and the quantity $VB^2$ over the last 14 years. These are plotted in Figure \ref{fig:Solar_Cycle_23_L_vs_R_vs_Kp_vs_CMEs}; similar plots have been made for the previous two cycles and the results are consistent, however solar cycle 23 has the best data coverage and so is the example used here. The data are split into \textbf{minimum} ($\bullet$) and \textbf{maximum} ($\times$) activity phases of the cycle, where \textbf{maximum} is defined as the \textbf{quarters with an average sunspot number greater than 40}. In order to compare correlation length with a consistent set of CME rates over the solar cycle, we take the same approach as \cite{Owens08}, and combine the early duty-cycle corrected LASCO CME rates \cite{StCyr2000}, processed LASCO CME catalogue rates \cite{Yashiro04, Gopalswamy08} (See also \url{http://cdaw.gsfc.nasa.gov/CME_list/index.html}), and the recent duty-cycle corrected STEREO Cor1 CME rates made by St Cyr and Xie, available at \url{http://cor1.gsfc.nasa.gov/}. See \cite{Owens08} for more detail.
\par

The top left panel of Figure \ref{fig:Solar_Cycle_23_L_vs_R_vs_Kp_vs_CMEs} plots the sunspot number against $\lambda_{|B|}$. There is a weak trend showing increasing correlation length with sunspot number for the times defined as solar minimum, but the longest correlation lengths are found after sunspot maximum in the declining activity part of the cycle. Comparing $\lambda_{|B|}$ with CME rate in the top right panel of Figure \ref{fig:Solar_Cycle_23_L_vs_R_vs_Kp_vs_CMEs} gives very similar results to sunspot number, showing some relation but with considerable scatter. The bottom two panels of Figure \ref{fig:Solar_Cycle_23_L_vs_R_vs_Kp_vs_CMEs} plot $\lambda_{|B|}$ against daily total magnetospheric activity index Kp and the calculated solar wind quantity $VB^2$, which is the amount of magnetic energy passing the spacecraft in a given time and a known driver for magnetospheric activity \cite{Akasofu79}. Both show a linear agreement with the measured $\lambda_{|B|}$ and reduced scatter compared to the CME rate and sunspot number. This supports the idea that $\lambda_{|B|}$ is related to the energy containing scale of the solar wind and that it is these $|B|$ scales that are geo-effective. Furthermore, this suggests that Kp or other magnetospheric activity indices could possibly be used as historical proxies for correlation in the solar wind magnetic field as well as its strength. 
\par
\begin{figure}
\centerline{\includegraphics[width=0.75\textwidth]{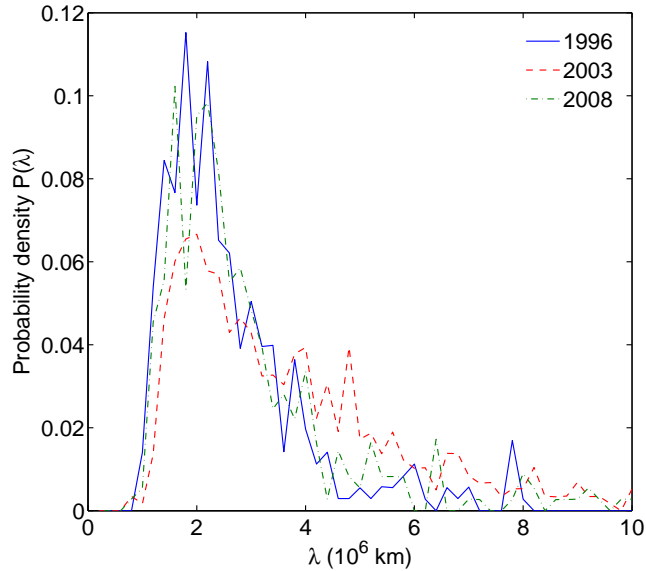}}
\caption{Probability density functions of daily calculated correlation lengths from three years of the OMNI data set, two \textbf{years with small correlation lengths close to solar minimum} (1996, solid line and 2008, dot-dashed line) and \textbf{the year with the largest average correlation length, close to but not directly coinciding with solar maximum} (2003, dashed line).}
\label{fig:PDF_Ls}
\end{figure}

Finally we calculate the probability distribution of daily estimates of $\lambda_{|B|}$ during the two years with lowest average $\lambda_{|B|}$ (1996 and 2008) and the year with the highest $\lambda_{|B|}$ (2003) in the OMNI data set. These are plotted in Figure \ref{fig:PDF_Ls}. The daily correlation lengths are calculated by fitting the same exponential function to each day of data rather than to quarterly averages: the resulting correlation lengths are very variable but their distribution provides insight into what processes are causing the long term variation. The two minimum years have similar shaped curves with a peak at approximately $1.8 \times 10^6$ km and a slight tail towards longer correlation lengths. 2003 has a peak at a similar length to the minimum years but it is much lower and the tail of longer correlation lengths is much increased.

\par

\section{Conclusions}

The averaged correlation lengths used here can be thought of as the average structure size in the variable being analysed. The values calculated here are consistent with previous estimates \cite{Matthaeus05, Wicks09}. As such these results can be explained by the continual presence of turbulent structures with a constant average size of approximately $1.5 \times 10^6$ km throughout the solar cycle. This provides the minimum correlation length for $|B|$ and the peak in Figure \ref{fig:PDF_Ls}, as well as the constant correlation length of the components of \textbf{B} and \textbf{V}. This value is unaffected by the solar cycle and is consistent with the continual presence of developed Alfv\'{e}nic turbulence at 1 AU. The correlation length of $|B|$ has an additional component added by solar activity causing it to change with the solar cycle, this could be partially caused by CMEs (Figure \ref{fig:Solar_Cycle_23_L_vs_R_vs_Kp_vs_CMEs}) but the highest correlation is found between $\lambda_{|B|}$ and $VB^2$, which is not necessarily related to any single type of event in the solar wind.
\par
The correlation of $\lambda_{|B|}$ with Kp and $VB^2$ suggests that the former is an estimate of the energy containing structure size in the solar wind. This could suggest that $\lambda_{|B|}$ is related to the average size of packets of plasma in the solar wind which would provide a physical outer scale in slow wind in particular. $\lambda_{|B|}$ is also consistent with the idea described in \cite{Borovsky08} of large scale structures or flux tubes in $|B|$ which contain, or at least provide an upper scale for, the turbulent solar wind. This also implies that historical records of magnetospheric activity could be used to infer average structure size in the solar wind as well as the magnetic field strength and solar wind velocity \cite{Lockwood99, Svalgaard07}.
\par
We note that $\lambda_{|V|}$ has reached an unprecedentedly large value in recent years and that this might be connected with the current decrease in solar wind density and ram pressure \cite{McComas08}, although we currently have no explanation for this. $\lambda_{|B|}$ had not yet returned to its previous minimum value by the end of 2008 suggesting that solar cycle 23 had not reached minimum by this time.

%
\begin{acks}
We would like to thank the NASA National Space Science Data Center (NSSDC) and Space Physics Data Facility (SPDF) for the OMNI, ISEE3 and IMP8 data sets, and the Principal Investigators Dr. Adam Szabo (GSFC) and Dr. Alan Lazarus (MIT) for data usage from the IMP8 spacecraft. This work was funded by the Science and Technology Facilities Council (STFC).
\end{acks}

%
%
%
%
%
%

\end{article}


%
%

\end{document}